\documentclass[11pt]{article}

\usepackage{latexsym}
\usepackage{amssymb}
\usepackage{amsmath}
\usepackage[arXiv]{optional} 
\opt{for_Paolo}{

}

\def\Cl{{\cal C}\ell}
\def\End{\textrm{End}}

\begin{document}
\thispagestyle{empty}
\begin{center}
\opt{for_Paolo,std}{
\null\vspace{-1cm}
{\footnotesize Available at:
{\tt http://publications.ictp.it}}\hfill IC/2009/011\\
\vspace{1cm}
United Nations Educational, Scientific and Cultural Organization\\
and\\
International Atomic Energy Agency\\
\medskip
THE ABDUS SALAM INTERNATIONAL CENTRE FOR THEORETICAL PHYSICS\\
\vspace{2.5cm}
}
{\bf FROM PURE SPINORS TO QUANTUM PHYSICS%
\opt{for_Paolo,std}{
\\
}
AND TO SOME CLASSICAL FIELD EQUATIONS LIKE%
\opt{for_Paolo,std}{
\\
}
MAXWELL'S AND GRAVITATIONAL}
\opt{for_Paolo,std}{
\\
\vspace{2cm}
}
\opt{arXiv}{
\medskip
}
Paolo Budinich\footnote{fit@ictp.it}\\
{\it International School for Advanced Studies (SISSA), Trieste, Italy%
\opt{for_Paolo,std}{
\\
}
and\\
The Abdus Salam International Centre for Theoretical Physics, Trieste, Italy.}\\
\opt{for_Paolo,std}{
\vfill
MIRAMARE -- TRIESTE\\
\medskip
}
\opt{for_Paolo}{
\bigskip
\Huge
27 November 2009
}
\opt{std}{
May 2009\\
\vspace{2cm}
}
\end{center}

\begin{abstract}
\opt{for_Paolo}{
\huge
\begin{center}
{\bf Abstract}
\smallskip
\end{center}
}
In a previous paper~\cite{PBudinich_2008} we proposed a purely mathematical way to quantum mechanics based on Cartan's simple spinors in their most elementary form of 2 component spinors. Here we proceed along that path proposing, this time, a symmetric tensor, quadrilinear in simple spinors, as a candidate for the symmetric tensor of general relativity. This is allowed now, after the discovery of the electro--weak model and its introduction in the Standard Model with $SU(2)_L$. 

The procedure resembles closely that in which one builds bilinearly from simple spinors an antisymmetric ``electromagnetic tensor'', from which easily descend Maxwell's equations and the photon can be seen as a bilinear combination of neutrinos. Here Lorentzian spaces result compact, building up spheres, where hopefully some of the problems of the Standard Model could be solved as pointed out in the conclusions.
\end{abstract}

\opt{std}{
\vfill

\newpage
}
\baselineskip=24pt
\setcounter{footnote}{0}
\opt{for_Paolo}{
\huge
}
\section{Introduction}

In a previous paper~\cite{PBudinich_2008} it was shown how, from the simplest, non trivial, two component spinors, which are pure, we may bilinearly and quite naturally obtain null vectors which are building up Minkowski momentum space. With these, in fact, one may formulate the well-known Weyl equations of motion for massless neutrinos as follows:\footnote{Neutrinos must be massless in order to obtain parity violations in the so-called weak interactions, like neutron decay.}
\begin{equation}
p_\mu\gamma^\mu (1+\gamma_5)\psi =0
\end{equation}
where $\psi$ is a spinor (Dirac) associated with the Clifford algebra $\Cl (3,1)$ with $\gamma_\mu$ its generators, and $\gamma_5$ its volume element.

In \cite{PBudinich_2008} $p_\mu$ are bilinearly obtained precisely from those two component pure spinors from which we started.

This was a first elementary result of Ref.~\cite{PBudinich_2008} which was aimed at searching a purely mathematical formulation of quantum mechanics. A similar way was followed by Einstein, Poincar\'e, and other outstanding mathematicians in the early part of last century, when they arrived at special and general relativity. In this way one could avoid the plague of paradoxes which instead, up to our time, ruined the development of quantum mechanics mostly deriving from the naive postulate that the Schr\"odinger (or Dirac) wave function $\Psi(x)$ has to collapse in the point $x=0$ if the particle it represents is observed and that is at the origin of many paradoxes (like the famous Einstein's bed and Schr\"odinger's cat).
In Ref.~\cite{PBudinich_2008} two more main results were obtained:

\noindent{\bf I.}\ while the classical dynamics of macroscopic bodies has to be formulated and dealt with in ordinary (Minkowski possibly curved) space-time, as we learned from Newton, Lagrange, Hamilton, Jacobi, Einstein, Poincar\'e and others, so here the basic geometry is an Euclid's one, including the concept of point-event, which is necessary to represent the center of mass of the macroscopic bodies while running along their trajectories or orbits (like the Kepler orbits).

\noindent{\bf II.}\ For atomic physics instead, the appropriate space for the formulation of dynamical quantum equations and for their solution is momentum-space bilinearly constructed from pure spinors, as it appears in (1) the first elementary example of the Weyl equations.\\

\section{Pure Spinors}

\'E. Cartan, the discoverer of spinors, underlined the great elegance of the spinors he named simple~\cite{Cartan_1937} (renamed pure by Chevalley~\cite{Chevalley_1954}).

He started by considering the correlations between spinors and totally null planes as follows: let $W = C^{2n}$ represent a complex Euclidean, $2n$ dimensional space, with well-defined quadratic forms. Then we may define the corresponding Clifford algebra $\Cl (2n)$ which, as known, may be considered as the endomorphism algebra in a $2^n$ dimensional space $S$ of spinors, and we write:
\begin{equation}
\Cl (2n) \cong \End\ S
\end{equation}
A spinor $\psi$ may be defined by the equation
\begin{equation}
Z_a\gamma^a\psi = 0;\quad a=1,2,\dots ,2n
\end{equation}
where $Z$ is a vector of $W$ (referred to a Cartesian orthonormal coordinate system) and $\gamma_a$, called the generators of $\Cl (2n)$, represent the univectors in the direction of the $2n$ orthonormal coordinates of $W$.

The generators $\gamma_a$ obey the anticommutation relations.
\begin{equation}
(\gamma_a\gamma_b+\gamma_b\gamma_a) = 2\delta_{ab}
\end{equation}

Let us now multiply equation (3) from the left by $\gamma^a Z^1_a$ and we obtain (because of equation (4)):
\begin{equation}
Z Z^1 \psi =0
\end{equation}
and, if $Z^1=Z$:
\begin{equation}
Z^2 \psi =0
\end{equation}
which means that the non zero spinor $\psi$, satisfying equation (3) (sometimes called Dirac spinor) defines a subspace of $W$ whose vectors are all null and/or mutually orthogonal, it is called the totally null plane associated to $\psi$ and indicated with $T_d (\psi)$ where $d$ is its dimension. It is known, and easy to prove, that the maximal possible value of $d$ is $n$, that is, one half of the dimension $2n$ of the space $W$.

In order to arrive at the Cartan definition of simple or pure spinors, let us define the ``volume element'' of $\Cl (2n)$:
\begin{equation}
\gamma_{2n+1}=\gamma_1\gamma_2\dots\gamma_{2n}
\end{equation}

It is easy to show that $\gamma_{2n+1}$ anticommutes with all the $\gamma_a$ and that $\gamma_1, \gamma_2,\dots, \gamma_{2n}, \gamma_{2n+1}$ generate the Clifford algebra $\Cl (2n+1)$ whose even subalgebra $\Cl_0 (2n+1)$ is isomorphic to the simple algebra $\Cl (2n)$:
\begin{equation}
\Cl_0 (2n+1) \cong \Cl (2n)
\end{equation}
With $\gamma_{2n+1}$ one may define what are called the Weyl spinors $\varphi^\pm_W$:
\begin{equation}
\varphi^\pm_W=\frac{1}{2}\left( 1\pm\gamma_{2n+1}\right)\psi
\end{equation}
where $\psi$ is the Dirac spinor defined by equation (3).

The Weyl spinors may be defined by the Weyl equations
\begin{equation}
Z_a\gamma^a\left(1\pm\gamma_{2n+1}\right)\varphi^\pm_W=0
\end{equation}
Obviously also $\varphi^\pm_W$ will define totally null planes in $W$. Well \'E. Cartan showed how a simple or pure spinor is isomorphic, up to a sign, to the maximal totally null plane of the Weyl spinor associated with a given Clifford algebra, and this property renders pure spinors complicated geometrical objects to deal with. In fact, while the dimension of the maximal totally null planes increase linearly with $n$, that of the spinors, increase with $n$ like $2^n$ and then high dimensional spinors which are used to explain some phenomena of elementary particle physics (up to $n=5:32$ component spinors), need a lot of (ten for 32 components) constraint equations in their components to render them pure. In short, all spinors up to $4$ components are pure (or equivalent to pure).

At present several people have tried to use pure spinors to explain both the high energy phenomena of the elementary particles and the unsolved problem of quantization of the gravitational field (through superstring theory) but with no acceptable results.

In fact, the so-called Standard model, now represented by the symbols $[U(1), SU(2)_L, SU(3)]$, which represents the symmetries presented in high-energy elementary particle phenomena, contains more than 20 constants representing charges, masses and so on, which may not be computed deriving them from acceptable theory.

But now with the possible prescription found in \cite{PBudinich_2008} that quantum dynamical problems have to be dealt with in momentum-space rather than in space-time, and because of a theorem we recently discovered, the situation could drastically change.\\

\section{The theorem}

Consider two Weyl spinors: $\psi$ and $\phi$ associated with a Clifford algebra $\Cl (2n)$ and define the vector of the space $W=C^{2n}$, with components:
\begin{equation}
Z_a=\langle\psi ,\gamma_a\phi\rangle
\end{equation}
where $a = 1,2,\dots ,2n$. The vector $Z$ is null: $Z_a Z^a= 0$,
if and only if one of the two spinors, either $\psi$ or $\phi$, is pure.
The proof is in Ref. \cite{PBudinich_Trautman_1989}.

It is known that in $2n$ dimensional spaces with Lorentzian signatures like $V=R^{2n-1,1}$ or $R^{1,2n-1}$ the vector components will be real if of the form $\tilde\psi\gamma_\mu\psi$, where $\tilde\psi = \psi^+\gamma_0$ and $\psi^+$ means $\psi$ hermitian conjugate while $\gamma_0$ is the time-like generator.

Suppose now $\psi$ to be pure then we will have that:
\begin{equation}
P_aP^a=0;\quad a=1,2,\dots , 2n
\end{equation}
with real components, (12) may be written in the form:
\begin{equation}
\pm P_\mu P^\mu = M^2_n+P^2_5+P^2_6+\dots
\end{equation}
where $\mu = 0,1,2,3$; which, might arrive (in the r.h.s.) at $P^2_{10}$ (exploiting the isomorphism between Dirac and Weyl spinors doublets and conformal covariance \cite{PBudinich_2008}).

It is interesting to observe that in the momentum space, where according to the conclusion of Ref. \cite{PBudinich_2008}, we have to formulate the dynamical problems of quantum physics, pure spinors may bilinearly define in Lorentzian momentum spaces compact manifolds consisting of spheres (with Poincar\'e invariant radii). This might obviously encourage the hope that not only the mathematical origin of quantum jumps might finally find a simple geometrical explanation, (the auto-vibrations of spheres are discrete) but also that, hopefully, the transitory sickness of the otherwise beautiful Standard Model, manifested by the symptom of its more than 20 unexplained constants, may finally be cured.\\

\section{First applications and results}

It might sound strange, but it is true: the first convincing confirmation of the above predictions was found by V. Fock more than 70 years ago: in 1935~\cite{Fock_1935}. He dealt with the historical problem (the first one to deal with atomic quantization) of the H-atom stationary states.

At that time the problem was solved through the Schr\"odinger equation for the electron possible orbits in the field of the proton. Fock, anticipating the conclusions of \cite{PBudinich_2008}, formulated it in momentum space on a sphere $S_3$ conceived as the one point compactification of ordinary $3$ dimensional momentum space $R^3$ with the following integral equation:
\begin{equation}
\psi (u) =
\frac{\alpha}{V(S_3)}\ \frac{mc}{p_0} \int_{S_3}\ \frac{\psi (u')}{(u-u')^2}\ d^3u'
\end{equation}
where $V(S_3) = 2\pi^2$ is the volume of the $S_3$ unit sphere; $\alpha = e^2/\hbar c$ is the fine structure constant, $p_0$ a unit of momentum, $m$ the (reduced) electron mass and $u$ is a unit vector indicating a point on the unit sphere $S_3$, equation (14) is the Fock equation in adimensional form. Fock solved it through harmonic analysis and found that setting $p_0 = (2m\ E)^{1/2}$ he obtained for $E_n$, representing the eigen vibrations of the sphere $S_3$, the energy levels $E_n$ of the H-atom stationary states, which in turn explain the Balmer spectrum of the H-atom. But it was V. Fock himself to draw from equation (14) the fundamental discovery that an integral equation on the sphere $S_3$, like equation (14), has to forsee the symmetry $SO(4)$ of its solutions (including obviously their classical counterparts -- the solar planets -- which is {\em true}, as later underlined by the great W. Pauli, Nobel prize winner).

Now we may go one step further representing the H-atom with a quadruplet of spinors say (Proton, Neutron, Neutrino and Electron) and adopt equation (13) with $n=4$ \cite{PBudinich_2006}: since it is well known that spinors up to 4 components are pure (or equivalent to pure, and this equivalence is exploited here)\footnote{It is interesting to have obtained the necessity of the Minkowski signature of momentum space which came out obviously at the first step of Ref. \cite{PBudinich_2008} when we were dealing with two component pure spinors; for which $M^2=0$. But now we are with 4 component spinors! and equation (14$'$) is correct (since 2 component spinors correspond to $n=1$).}~\cite{PBudinich_2006}
$$ P_\mu P^\mu=M^2_4+P^2_5+P^2_6+P^2_7+P^2_8 \eqno(14')$$
from which we obtain $S_3$ with Poincar\'e invariant radius $M_4$ therefore we may expect that the spinorial $S_3$ may reflect the Poincar\'e invariance; and in the energy levels found by Fock we have the obviously relativistic form \cite{PBudinich_2006}:
\begin{equation}
E_n=\frac{\alpha^2}{2}\ \frac{mc^2}{n^2};\quad n=1,2,3\dots
\end{equation}

If we take this first result as an indication of the validity of the hypothesis formulated above \cite{PBudinich_2008} we may affirm that the first example of atomic quantization confirms that the H-atom discrete energy levels (generating its discrete optical spectrum -- Balmer series) are generated by the computed manifolds determined by pure spinors in momentum space. In particular by the eigen vibrations of those spheres we may also geometrically explain the quantum-jumps.

Should this result have confirmations by further examples then the historical way to arrive at the solution of the problem of the H-atom stationary states would not only be more complicated but also wrong. It would be wrong to consider the H-atom as a proton plus an orbitating point electron to be then substituted arbitrarily by wave functions.

As stated in \cite{PBudinich_2008} we have to formulate the quantum problems in momentum space where we have the mathematical possibility to integrate null vectors to generate strings \cite{PBudinich_1986}, which are non local objects and might be at the origin of the concept of wave function without giving origin to paradoxes, as will be further discussed, elsewhere.\\

\section{The possibility to cure the Standard Model}

V. Fock solved equation (14) computing the adimensional factors $mc/p_0$ through harmonic analysis from the ball $B_3$ to $S_3$. We tried (with P. Nurowski) to compute, through harmonic analysis, also the other adimensional factor $\alpha = e^2/\hbar c$ in equation (14) obtaining all the factors but one. In fact, $\alpha$ was computed for the first time by M.A. Wyler in 1969 \cite{Robertson_1971} with the result:
\begin{equation}
\frac{e^2}{\hbar c} = \frac{8\pi V(D_5)^\frac{1}{4}}{V(S_4)V(Q_5)} = (137,0608)^{-1}
\end{equation}
differing less than $1/10^6$ from the experimental value.

He used group theoretical methods, however declared (private communication) to be not familiar with spinor theory. There are more authors who with other methods computed other quantum constants (including $\alpha $) relevant for particle physics \cite{Gonzalez}. Now the point is if their computations may be correlated or derived from the compact manifolds in momentum space derived from pure spinors.\\

\section{The wave function}

If we follow the suggestion of Ref.\cite{PBudinich_2008} that quantum dynamics has to be formulated and dealt with in momentum space, then we have to abandon the concept of wave function representing the electron say, in the Schr\"odinger equation (at the source of several paradoxes). As we have seen in momentum space there is no possibility of defining the point event. We may instead define integrals of null vectors, bilinear in pure spinors which uniquely define strings, which are non local but relativistically covariant; furthermore, for their existence in nature, there is experimental evidence since they were discovered from the interpretation of some experiments performed at CERN (Geneve). It should not be difficult to obtain from relativistic non local objects also non relativistic ones like in the Fock integral equation, which, as seen above, presents relativistic properties.\\

\section{From the quantum to the classical field equations}

It is well known how, from the Weyl equation for massless neutrinos, equation (1) one may easily obtain the classical Maxwell's equations which, when quantized give the equations of motion for the massless and chargeless photon.
It also results from \cite{PBudinich_2008} that for $4$ dimensional momentum space $P_\mu P^\mu =0$; that is $M_4^2 = 0$.

In this way the massless photon may be conceived as bilinearly generated by massless neutrinos representing elementary pure spinors (but notoriously not their bound states).

Already \'E. Cartan named electromagnetic tensor the one contained in the Clifford algebra $\Cl (2n)$:
$$
F^{(\pm)}_{\mu\nu} = \tilde\psi \left[\gamma_\mu , \gamma_\nu\right] (1\pm\gamma_5) \psi
$$
In fact, if we apply the Weyl operator of equation (1) we obtain:
\begin{eqnarray}
p^\mu F^{(+)}_{\mu\nu} &=& 0\nonumber \\
p_\mu\varepsilon^{\mu\rho\tau\lambda} F^{(-)}_{\tau\lambda} &=& 0
\end{eqnarray}
which are the well known Maxwell's equations in vacuum.

At this point a natural question presents itself -- why not try to get from pure spinors also the gravitational field equations? These equations present enormous difficulties that for a long time blocked the progress of theoretical physics and the efforts of the best present physicists in Trieste (and no longer present, like my dear friend Dennis Shama). The best answer for a long time has been: because it is too difficult to get symmetric tensors from Clifford Algebras.

But now, there is a way to generate symmetric tensors from Clifford algebras: they are entered through the $SU(2)_L$ of the electroweak model (represented by $SU(2)_L$ in the standard model). And then let us try.

Let us then consider an element of the electroweak model of the form:
\begin{equation}
J_{\mu\nu}=\tilde\psi_1\gamma_\mu (1+\gamma_5)\psi_2\tilde\psi_3\gamma_\nu (1+\gamma_5)\psi_4
\end{equation}
it is obviously symmetric like should be the metric tensor (fundamental for general relativity!) and it is quadrilinear in pure spinors.

The equation for the massless and chargeless photon has been derived from Maxwell's equations.
A similar result should also be obtained for the massless and chargeless graviton, which however will have to result as quadrilinear of massless neutrinos since its spin is $2$ instead of $1$; and as we will see, this can be done in the following way, precisely starting from the standard model in its form:
$U(1) SU(2)_L SU (3)$ where $SU (2)_L$ represents the origin of the Electro Weak mode and contains products of the left-handed currents which, through equation (1) for the massless neutrinos, may produce the gravitational equation in flat space in a similar way as they produce Maxwell's equations for the massless photons.

In fact, suppose that in equation (18) $\psi_1\psi_2\psi_3\psi_4$ represents the existing weak decay, Neutron $\to$ Proton $+\ \tau$ lepton $+\ 2$ neutrinos (or the ones in which the fermions are substituted by one of the 3 twins of the same family), and we would get the graviton if we could consider chargeless the $(1+\gamma_5)$ proton and $(1+\gamma_5)$ lepton which appear on the right-hand side that should be both chargeless and massless like the two neutrinos. This possibility should certainly be discussed and analyzed further.

The situation appears difficult but not desperate since from the Weyl equation for the neutrino (1) we obtain the Maxwell's equations (17), notoriously extendable also to charges. From these in turn, we may derive the equations for the chargeless and massless photon. A similar procedure could be possible also for gravity since, as well known, its renormalizability was not solved through sophisticated supergravity and it is quite possible that it could be attacked going back to the first steps as suggested above.

A possible way could be: let us start from the obviously symmetric tensor $J_{\mu\nu}$ in (18) and postulate the obviously equivalent equations
$$
p^\mu J_{\mu\nu} = 0 = p^\nu J_{\mu\nu}
\eqno(19)
$$
which are similar to the first of the equations (17) from which one derives the zero mass of the photon. It remains to be discussed if and which phenomena are represented by the equations with the emisymmetric tensor $\varepsilon^{\mu\nu\rho\tau}$.

\section{Conclusions and possible follow-up}

In this paper we followed the suggestions of one of our great predecessors: \'Elie Cartan, the discoverer of spinors. One of the main Cartan's merits is to have underlined the importance of simple spinors and of being the first to see in Clifford algebras an emisymmetric tensor, that he named electromagnetic tensor. From this tensor one may derive the Maxwell's equations (17) from which only the first one, the simplest, is necessary to derive the zero mass and charge of the photon which, then, may be interpreted as bilinear in massless neutrinos (but not their bound state).

Another great physicist was P.A.M. Dirac who, in 1938 discovered the ``deep connection in Nature between cosmology and atomic physics'' \cite{Dirac_1938} used in \cite{PBudinich_2008}.
%


To get the graviton, one needs a symmetric tensor that can be built quadrilinearly in Weyl spinors. This construction could hopefully also provide suggestions on the quantization of the gravitational field which failed up to now despite several efforts (and whose history still waits to be written). %
Being in the frame of general relativity the appropriate field where to look for suggestions is probably Cosmology, as suggested in the paper, where certainly the space-time curvature must play a central role.

%
%
\opt{std, for_Paolo}{
\bigskip
}
\section*{Acknowledgments}

The author wishes to express his gratitude to M. Budinich for fundamental criticism and suggestions.

\opt{std, for_Paolo}{
\bigskip
}
\opt{arXiv}{
\newpage
}

\end{document}